\documentclass[12pt]{iopart}
\usepackage[left=1.5cm, right=1.5cm, top=1.785cm, bottom=2.0cm]{geometry}
\usepackage{graphicx} 
\usepackage{lastpage}
\usepackage{float}
\usepackage{fancyhdr}
\usepackage{fnpos}
\usepackage[english]{babel}
\addto{\captionsenglish}{%
  
}
\usepackage[T1]{fontenc}
\usepackage[usenames,dvipsnames]{xcolor}
\usepackage{setspace}
\usepackage[compact]{titlesec}
\usepackage{hyperref}
\usepackage{siunitx}
\newcommand{\FAU}{Institute of Micro- and Nanostructure Research (IMN) \& Center for Nanoanalysis and Electron Microscopy (CENEM), Friedrich Alexander-Universität Erlangen-Nürnberg, IZNF, 91058 Erlangen, Germany}
\newcommand{\UCBMaterials}{Department of Materials Science and Engineering, University of California Berkeley, Berkeley, California 94720, United States}

\newcommand{\LBLMSD}{Materials Sciences Division, Lawrence Berkeley National Laboratory, Berkeley, California 94720, United States}

\newcommand{\Foundry}{The Molecular Foundry, Lawrence Berkeley National Laboratory, Berkeley, California 94720, United
States}
\begin{document}

\title{High-resolution 3D phase-contrast imaging beyond the depth of field limit via ptychographic multi-slice electron tomography}

\author{Andrey Romanov}
\address{\FAU}
\author{Min Gee Cho}
\address{\UCBMaterials}
\author{Mary Cooper Scott}
\address{\UCBMaterials}
\address{\LBLMSD}
\address{\Foundry}
\author{Colin Ophus}
\address{\Foundry}
\author{Philipp Pelz}
\address{\FAU}
\ead{philipp.pelz@fau.de}
\vspace{10pt}
\begin{indented}
\item[]September 2023
\end{indented}

\begin{abstract}
Resolving single atoms in large-scale volumes has been a goal for atomic resolution microscopy for a long time. Electron microscopy has come close to this goal using a combination of advanced electron optics and computational imaging algorithms. However, atomic-resolution 3D imaging in volumes larger than the depth of field limit of the electron optics has so far been out of reach. Electron ptychography, a computational imaging method allowing to solve the multiple-scattering problem from position- and momentum-resolved measurements, provides the opportunity to surpass this limit. Here, we experimentally demonstrate atomic resolution three-dimensional phase-contrast imaging in a volume surpassing the depth of field limits using multi-slice ptychographic electron tomography. We reconstruct tilt-series 4D-STEM measurements of a $\mathrm{Co_3O_4}$ nanocube, yielding \SI{1.75}{\angstrom} resolution in a reconstructed volume of $\mathrm{(18.2 nm)^3}$.
\end{abstract}
\section{Introduction}
Aberration-corrected scanning transmission electron microscopy (STEM) is a highly potent technique for visualizing materials at the atomic level and analyzing their chemical makeup. Advances in equipment, insights into the interactions between electrons and matter, and enhancements in automation and data analysis have propelled its prominence for materials characterization. Lately, the application of computational imaging approaches that leverage 2D position- and 2D momentum-resolved measurements (4D-STEM) has emerged as a formidable method, enabling the imaging of heavy and light elements, pinpointing atomic positions with picometer-level accuracy, and charting out phase, orientation, and strain across extensive areas.
With the latest generation of ultrafast direct electron detectors, new capabilities and methods are unlocked that were previously inaccessible. Fast direct electron detectors' high detective quantum efficiency allows imaging of beam-sensitive materials at unprecedented resolution and precision. Computational imaging methods like integrated center of mass imaging (iCOM) \cite{Lazic_Bosch_Lazar_2016}, optimum bright-field (OBF) imaging \cite{Ooe_2023}, and electron ptychography \cite{Yang_2016} utilize 4D-STEM datasets to produce high-contrast, high-resolution images of beam-sensitive materials like metal-organic frameworks and zeolites \cite{Peng_2022,Sha_Cui_Li_Zhang_Yang_Li_Yu_2023}, Li-ion battery materials \cite{Lozano_Martinez_Jin_Nellist_Bruce_2018}, polymers and perovskite materials\cite{Reis_2018}. A new frontier of electron microscopy is the recovery of three-dimensional information from single 4D-STEM datasets. The first results demonstrated three-dimensional optical sectioning of weakly scattering nanomaterials from a single 4D-STEM dataset using electron ptychography \cite{Gao_2017_3dptycho}. More recent work employing advanced multi-slice ptychographic reconstruction algorithms allowed the solution of multiple scattering in thick crystalline slabs and reached the physical resolution limits in the transverse direction \cite{chen2021electron}. Additionally, this method can surpass the axial resolution limit set by the numerical aperture of the electron optics. So far, axial resolutions of several nm have been demonstrated using this method \cite{chen2021electron}. Multi-slice ptychography has since been applied to a wealth of materials. This quick adaptation of multi-slice ptychography to study different materials systems demonstrates the usefulness of 3D phase-contrast imaging based on 4D-STEM. However, the axial resolution of multi-slice ptychography is currently limited to around 3nm, and sub-nm resolution can only be expected with extremely high fluence and the most advanced electron optics \cite{Chen_Shao_Jiang_Muller_2021}.\\
A solution to this problem of limited axial resolution is the collection of tilt-series 4D-STEM data and subsequent tomographic reconstruction. This approach yields 3D atomic resolution phase-contrast but comes with the additional complexity of aligning atomic-resolution high-dimensional datasets. Atomic resolution ptychographic tomography was recently demonstrated on a 2D material and on a complex double-wall carbon nanotube filled with a $\mathrm{Zr_{11}Te_{50}}$ structure.\\
These works used the projection approximation to reconstruct 2D projection images of the sample, from which the 3D atomic structure was either directly determined by peak fitting, or a 3D volume was reconstructed from which an atomic model was determined. The projection approximation limits the applicability of this approach to a few nm thick materials.\\
Here, we significantly advance three-dimensional imaging by recovering a three-dimensional phase-contrast volume 3 times larger than the depth of field at a resolution that allows us to distinguish single atoms in three dimensions. We demonstrate multi-slice ptychographic electron tomography with $\mathrm{Co_3O_4}$ nanocube supported on an amorphous carbon substrate in a volume with \SI{18.2}{\nano\meter} side length.
\section{Related work}
Since many advances in electron ptychography were first demonstrated with photons at visible and X-ray wavelengths, it is highly instructive for further method development in STEM phase-contrast imaging to review and categorize the algorithmic and experimental techniques developed over the last decade in ptychography with photons. The foundation for later tomographic algorithms extending beyond the depth of field limit was laid in 2012 with the realization of multi-slice ptychography \cite{maiden2012ptychographic} at visible wavelengths. Later, this algorithm was applied also in the hard X-ray regime \cite{Tsai_Sicairos_2016}, \cite{hu2022multi}, but so far the application has been limited to proof-of-principle demonstrations, mostly because the depth of field of X-ray optics is quite large and a multi-slice treatment was often not necessary. The first demonstration of a 2-step approach to ptychographic tomography showed a 3D reconstruction of bone at X nm resolution using hard X-rays and zone-plate optics \cite{Dierolf_2010}. This approach is now prevalent at many X-ray beamlines worldwide, offering improved resolution and sensitivity compared to other phase-contrast imaging methods. The traditional two-step approach can be significantly improved by reconstructing the volume directly from the diffraction measurements in a joint reconstruction using the single-slice approximation \cite{Chang_Enfedaque_Marchesini_2019}\cite{kahnt2019coupled}\cite{aslan2019joint}\cite{nikitin2019photon}. The next step in model complexity is to solve a multi-slice ptychography problem at every tilt-angle and then using this information in a subsequent tomography reconstruction. This approach has been realized at visible \cite{li2018multi} and x-ray \cite{kahnt2021multi} wavelengths in proof-of-principle experiments. The multi-slice ptychographic and tomographic reconstructions can also be combined in a joint multi-slice ptychographic tomography reconstruction algorithms, that couples all diffraction measurements directly to the reconstructed volume \cite{Gilles_Nashed_Du_Jacobsen_Wild_2018}, \cite{ozturk2018multi} \cite{jacobsen2018relaxation} \cite{huang2019resolving}\cite{Du_Nashed_Kandel_Gürsoy_Jacobsen_2020}. This approach requires good initial guesses of all nuisance parameters and has not been realized yet experimentally. Another interesting experimental development is precession multi-slice ptychography
\cite{Shimomura_Hirose_Takahashi_2018}, where the sample is tilted to a few small angles, and a joint 3D reconstruction is performed from these angles with increased axial resolution. This approach might be desirable in STEM since precession can be achieved very quickly with electron optics.

In the transmission electron microscope, multi-slice ptychography was first demonstrated five years after the visible light experiments \cite{Gao_2017_3dptycho}, yielding Angstrom-scale transverse resolution and tens of nm depth resolution of carbon nanotubes. Recently, sub-Angstrom resolution multi-slice electron ptychography was demonstrated using a highly efficient high-dynamic range electron detector. The technique has since seen rapid adoption by electron microscopy, resulting in high-resolution imaging of zeolites \cite{zhang2023three}, three-dimensional imaging of dislocation cores in $SrTiO_3$ \cite{sha2023sub}, and computational crystal mistilt correction \cite{sha2022deep}. Ptychographic electron tomography was first proposed at atomic resolution using a multi-slice ptycho-tomography algorithm \cite{vandenbroek_Koch_2013}, a relatively small tilt range of 10 degrees, and a low electron energy of 40keV to enhance multiple scattering. In parallel, ADF-STEM-based atomic resolution tomography \cite{Scott_2012} was developed as a technique to solve the atomic structure of small nanoparticles and 2D materials \cite{Miao_Ercius_Billinge_2016}.
Ptychographic electron tomography was first realized at the nanoscale \cite{Ding_2022}, demonstrating improved contrast for imaging of hybrid materials. Another approach to 3D phase-contrast imaging with ptychography used a single-particle approach to reconstruct the 3D volume of a virus particle from many copies \cite{Pei_Zhou_Huang_Boyce_Kim_Liberti_Hu_Sasaki_Nellist_Zhang_2023}.
As a result of the success of ADF-STEM-based atomic resolution tomography, Chang et al. proposed a two-step method for ptychographic atomic tomography \cite{chang2020ptychographic}, showing improved contrast for weakly scattering elements. This approach was recently realized experimentally, using a fast-framing direct electron detector \cite{Pelz_2022}. Hofer et al. recently demonstrated a few-tilt ptycho-tomography two-step approach tailored to solving the 3D structure of 2D materials \cite{Hofer_2023}.
Joint single-slice ptycho-tomography for STEM has neither been proposed nor demonstrated yet, although they could yield significant improvement for nanoscale phase-contrast imaging, where multiple scattering effects are less pronounced, and missing-wedge problems persist. Joint multi-slice electron ptycho-tomography has been considered again recently, this time using the entire tilt range available in the TEM and compared with single-slice reconstruction \cite{Lee_Lee_Park_Ophus_Yang_2023}.
Using the joint reconstruction algorithm, a significant improvement in reconstruction quality over single-slice or two-step approaches is visible. This improvement may warrant the higher difficulty of reconstructing experimental data with these methods. From an experimental point of view, the question of how all parameters of such a joint reconstruction can be initialized with a sufficiently good initial guess is still unsolved. Here, we see the potential of the two-step multi-slice ptychographic tomography approach, which yields initial estimates for all tilt-dependent probes, subpixel probe positions, and initial Euler angles. The results of the two-step approach that we demonstrate here could then be used to bootstrap a joint reconstruction.
\section{Results}
\subsection{Experiment}
\begin{figure*}[ht!]
\includegraphics[width=1\textwidth]{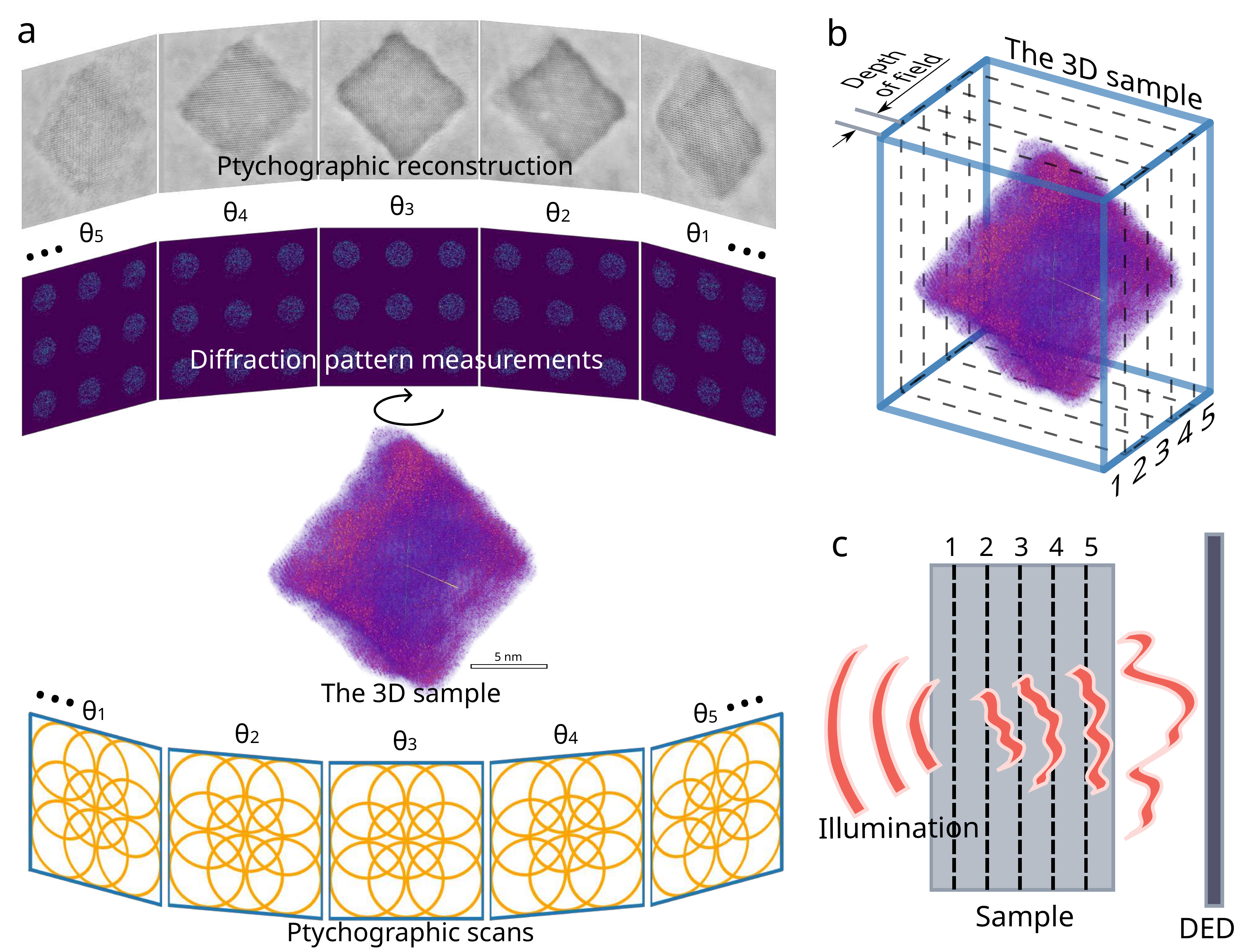}
\caption{\label{fig:figure1} Atomic resolution multi-slice ptychographic electron tomography. a) Scheme of the reconstruction algorithm and tilt-series 4D-STEM measurements. b) schematic of the slicing of the volume along the depth direction c) }
\end{figure*}

Figure \ref{fig:figure1} shows the overall scheme of the two-step approach for multi-slice ptychographic reconstruction. At each tilt angle, a 4D-STEM scan is taken with overlapping probes as indicated n Fig. \ref{fig:figure1} a). 
A tomographic tilt series was acquired from the $\mathrm{Co_{3}O_{4}}$ nanocube using the TEAM 0.5 microscope and TEAM stage \cite{Ercius_Boese_Duden_Dahmen_2012} at the National Center for Electron Microscopy in the Molecular Foundry. We recorded four dimensional-scanning transmission electron microscopy (4D-STEM) datasets \cite{ophus2019four} with full diffraction patterns over\num{800}x\num{800} probe positions at each tilt angle. The diffraction pattern images were acquired with the 4D Camera prototype, in-house developed in collaboration with Gatan Inc., a direct electron detector with \num{576}x\num{576} pixels and a frame rate of \SI{87}{\kilo\hertz} \cite{ercius20204d}, at 200 kV in STEM mode with a \SI{23}{\milli\radian} convergence semi-angle, a beam current of \SI{70}{\pico\ampere}, estimated from the flu-screen measurement, a real-space pixel size of \SI{0.33}{\angstrom}, and camera reciprocal space sampling of \SI{173.6}{\micro\radian} per pixel. These settings amounted to an accumulated fluence of \SI{4.6e4}{\elementarycharge\per\angstrom^2} per projection and \SI{1.6e6}{\elementarycharge\per\angstrom^2} for the whole tilt series. The tilt series was collected at 36 angles with a tilt range of +\num{63} to \num{-58} degrees. To minimize the total electron exposure, focusing was performed at a resolution of 80 kX before switching to high magnification for data collection.
\subsection{Multi-slice ptychography reconstruction}
\begin{figure*}[ht!]
\includegraphics[width=1\textwidth]{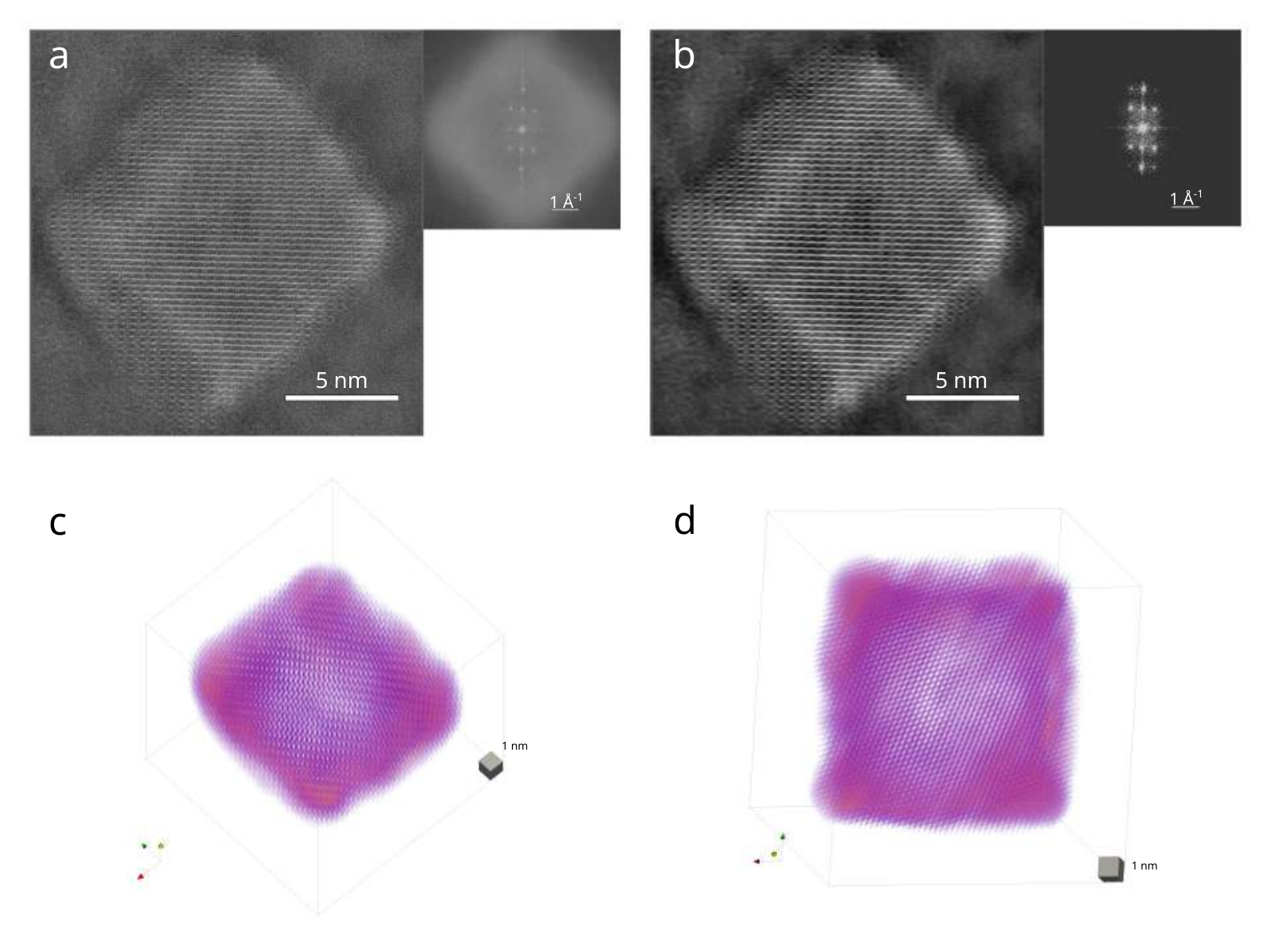}
\caption{\label{fig:figure2} Exemplary ptychographic reconstruction a) before and b) after denoising. c) Reconstructed volume, oriented along the <11-2> direction. d) reconstructed volume viewed along the <228> direction.}
\end{figure*}
For each 4D-STEM scan, we perform a multi-slice ptychography reconstruction using the implementation by Sha et al. \cite{sha2022deep}. Exemplary reconstructed summed slices of the phase-contrast images are shown in the top panel of \ref{fig:figure1} a). We then perform a tomographic reconstruction with joint Euler angle refinement and subpixel alignment as in \cite{Pelz_2022}.
\subsection{Tomography reconstruction}
\begin{figure*}[ht!]
\includegraphics[width=1\textwidth]{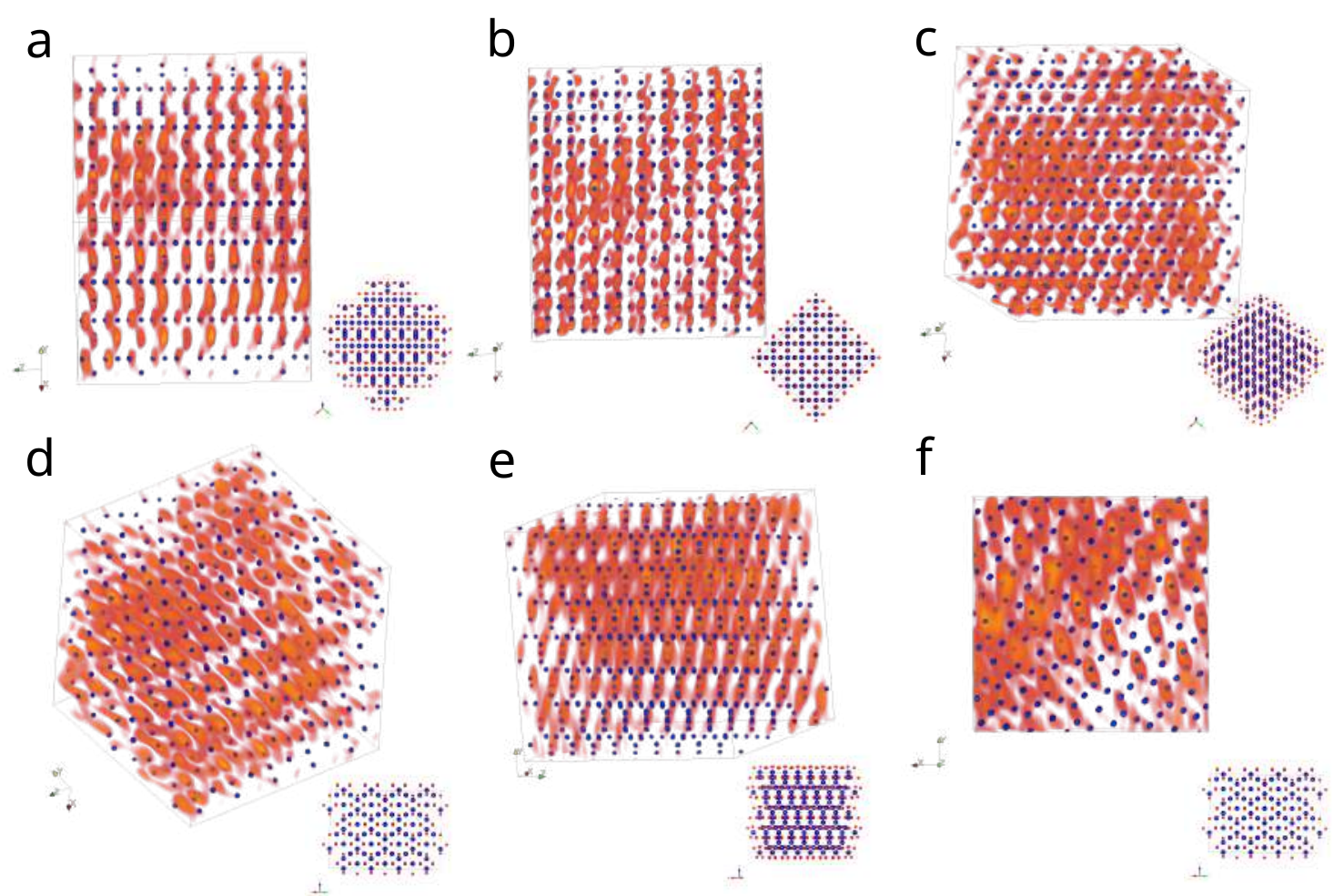}
\caption{\label{fig:figure3} Central subvolume of size \num{2.8} x \num{2.8} x \num{2.8} $\mathrm{nm}^3$ viewed from different directions in orthographic projection. a) <11-2>, b) <001>, c) <228>, d) <110>, e) <130>, f) <110>. Insets: atomic model of $\mathrm{Co_{3}O_{4}}$ viewed the corresponding direction.}
\end{figure*}
From the initially-aligned tilt series, we then perform a 3D reconstruction using joint reconstruction and rigid alignment as in \cite{Pelz_2022}, implemented with the automatic gradient calculation of the pytorch package \cite{paszke2019pytorch} using the projections obtained from multi-slice ptychography. \\
To refine tilt series alignment, we've devised a two-step optimization algorithm aimed at 3D reconstruction from down-scaled resulted projection images. In the initial step, we reconstruct the volume with a fixed alignment, and in the subsequent step, we refine the alignment with a fixed resulting volume. We executed this algorithm first on a 10\% resolution of the original ptychographic images for translation optimization and then on 20\% of the original for rotation optimization. Furthermore, we applied a 40\% intensity threshold to the original projections before down-scaling to mitigate noise and accentuate the contours of the nanocube.\\
Further, we utilized the resulting down-scaled reconstructed volume to build a full-scale mask volume by binarization after a certain value and following blurring, which quite accurately describes the outline of the nanocube. The projection images derived from this mask volume are then applied to the original ptychographic images to mask out substrate intensity outside of the cube. This approach compels a subsequent full-scale reconstruction to focus more intensively on the inner structure of the nanocube.\\
We conducted the final reconstruction with additional implementation of Gaussian blurring with $\sigma=$\SI{0.5} pixel of reconstructed volume to avoid populating frequencies beyond Nyquist. The resulting volume is shown in Fig. \ref{fig:figure3} c) and d). Since the $\mathrm{Co_{3}O_{4}}$ cube has a side length of about \SI{13}{\nano\meter}, we reconstruct a total of five slices along the depth of the cube.
\subsection{Analysis}
Due to the lack of background subtraction, the reconstructed volume exhibits higher noise than volumes of metal nanoparticles reconstructed previously ADF-STEM tomography, where the background noise of the substrate is less prominent and can be subtracted as a preprocessing step. Since the volume displays slowly varing contrast for the Co atomic peaks, instead of tracing atomic peaks one by one as done in previous AET work, we register the experimental Co lattice with the crystallographic model of $\mathrm{Co_{3}O_{4}}$ by first finding roughly the right rotation manually and then iteratively registering and rotating the experimental volume to optimize the coincidence. This gives us expected atomic sites of Co and O atoms, which we then extract from the volume and translationally align to create average atom volumes for resolution estimates. The extraction of the O atomic sites and averaging yielded no distinct peak in the average O volume, which indicates that the sensitivity of the reconstruction is not high enough on average to identify single O atoms.
We estimate the resolution of the Co atoms similar to \cite{chen2021electron}. First, we extract the mean Co atomic volumes from the experimental reconstruction and upsample to \SI{0.1}{\angstrom} pixel size. Then we simulate the Co atomic potential with the abTEM \cite{madsen2021abtem} package by averaging 250 frozen phonons at \SI{293}{\kelvin}, with standard deviation $\sigma=$\SI{0.075}{\angstrom} calculated from crystalline Fe \cite{sears1991debye} and convert them to the 3D transmission function. Subsequently, we solve an optimization problem that minimizes the difference between a Gaussian PSF-blurred transmission function and the mean experimental atom volume for Co atoms. The FWHM of the determined PSF follows approximately the Abbe resolution \cite{chen2021electron}. For the mean Co atom, we determine a resolution of $\mathrm{d_{\perp_x}}=$\SI{1.19}{\angstrom} and $\mathrm{d_{\perp_y}}=$\SI{1.68}{\angstrom} perpendicular to the missing wedge and $\mathrm{d_{\parallel}}=$\SI{2.39}{\angstrom} along the missing wedge direction, giving a mean 3D resolution of $\mathrm{d_{Co}}=$\SI{1.75}{\angstrom}. 
\section{Discussion}
We have demonstrated 3D phase-contrast imaging at a resolution of \SI{1.75}{\angstrom} in a reconstructed volume of $\mathrm{(18.2 nm)^3}$, succeeding in surpassing the depth-of-field limit of high-resolution electron microscopy by a factor of 3. Due to the noise level of the reconstruction, we could not reliably resolve the oxygen atoms in the $\mathrm{Co_{3}O_{4}}$ lattice. We attribute the noise level to the amorphous substrate, which resulted in significant background at high spatial frequencies. This background cannot simply be subtracted as in ADF-STEM tomography. Since the first demonstrations of ptychographic electron tomography reconstructed thin objects over vacuum, and most published results on atomic resolution multi-slice ptychography use a thin-film geometry without substrate, this raises the question of whether ptychographic electron tomography is a promising method for reconstructing single light atoms in samples that need to be suspended over substrates. We think there is enough room for improvement both on the algorithmic aspects, including joint reconstruction algorithms, and on the hardware side by increasing the detection efficiency of the fast-framing 4D-STEM detectors such that this question cannot be answered definitively in this article. Certainly, ptychographic electron tomography has an advantage over the ADF-STEM-based method when it comes to experimental automation since precise focusing is not required for ptychography, making the experiments less challenging, although more data-intensive, and indicating a bright future for 4D-STEM-based tomography at atomic resolution.

\section{Author Contributions}
A.R.: Methodology, Investigation, Writing - original draft, Visualization\\
P.M.P: Conceptualization, Investigation, Writing - original draft, Writing - Review \& Editing, Funding acquisition, Project administration\\
C.O.: Investigation, Writing - Review \& Editing\\
M.C.S: Investigation, Writing - Review \& Editing, Funding acquisition\\
M.C.: Investigation, Writing - Review \& Editing\\
P.M.P: Conceptualization, Investigation, Writing - original draft, Writing - Review \& Editing, Funding acquisition, Project administration\\
\section{Acknowledgements}
PP is supported by the Strobe STC research center, Grant No. DMR 1548924 and by an EAM starting grant project ScatterEM.
MCS is supported by the Strobe STC research center, Grant No. DMR 1548924.
A.R. is supported by an EAM Starting grant project ScatterEM.
CO acknowledges support from the Department of Energy Early Career Research Award program. Work at the Molecular Foundry was supported by the Office of Science, Office of Basic Energy Sciences, of the U.S. Department of Energy under Contract No. DE-AC02-05CH11231. 
PP gratefully acknowledges the scientific support and HPC resources provided by the Erlangen National High Performance Computing Center (NHR@FAU) of the Friedrich-Alexander-Universität Erlangen-Nürnberg (FAU) under the NHR project AtomicTomo3D. NHR funding is provided by federal and Bavarian state authorities. NHR@FAU hardware is partially funded by the German Research Foundation (DFG) – 440719683.
The 4D Camera was developed under the DOE BES Accelerator and Detector Research Program, with collaboration from Gatan, Inc. 
We thank Peter Ercius for help with 4Dcamera data acquisition.
We would like to thank Gatan, Inc. as well as P. Denes, A. Minor, J. Ciston, J. Joseph, and I. Johnson who contributed to the development of the 4D Camera.
\section{References}
\bibliographystyle{unsrt}
\bibliography{main}
\section{Supplementary Material}
\section*{Ptychographic tilt series}
\begin{figure*}[ht!]
\includegraphics[width=1\textwidth]{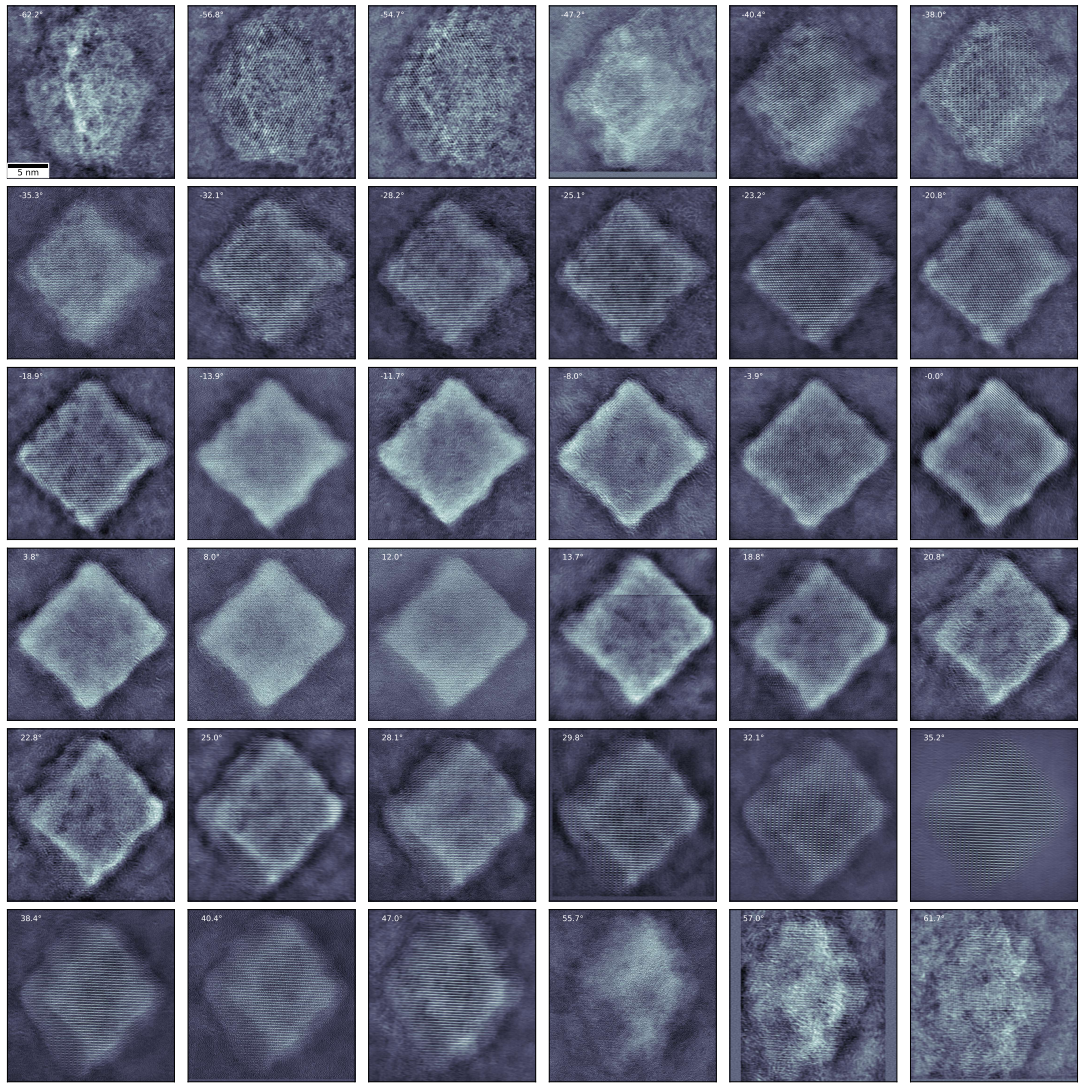}
\caption{\label{fig:tilt_series} Ptychographic multi-slice electron tomography tilt series of the nanoube. The 36 phase-contrast projection images with a tilt range from +\num{63} to \num{-58} degrees (shown at top left of each panel) were measured with the 4D Camera and reconstructed using the LSQML algorithm \cite{sha2023sub}. The projections are denoised with the BM3D algorithm.}
\end{figure*}
\newpage
\pagebreak
\section*{Reconstructed probes}
\begin{figure*}[ht!]
    \includegraphics[width=1\textwidth]{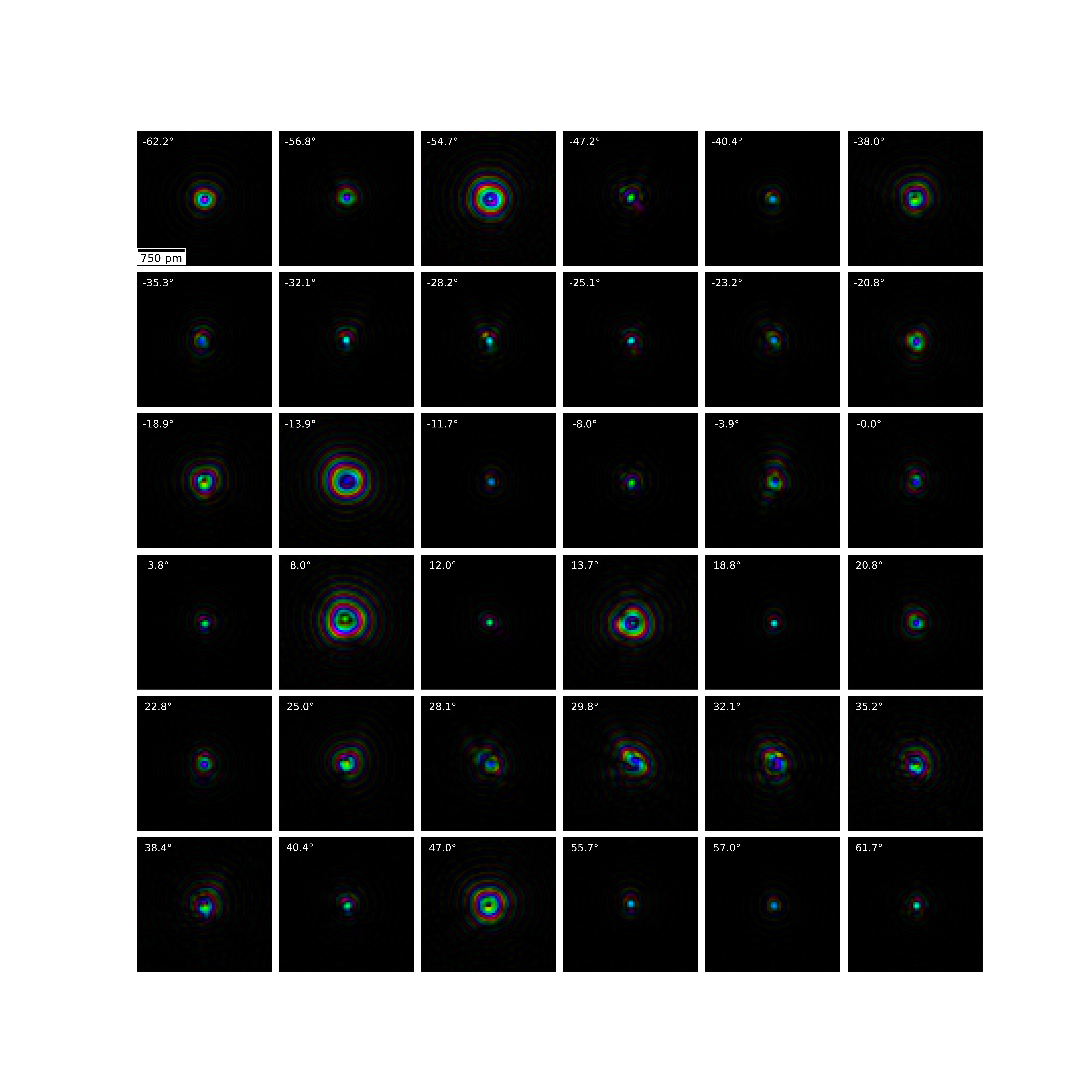}
    \caption{\label{fig:probes} Reconstructed first probe mode wave function for each tilt in real space. Amplitude is shown using color saturation, while phase of each pixel is given by the hue.}
\end{figure*}
\newpage
\pagebreak
\section*{Ptychographic reconstruction parameters}
\begin{table}[ht!]
\begin{tabular}{ l l}
$\texttt{Num\_probe\_state}$ & 4\\  
$\texttt{Num\_slice}$ & 5\\  
$\texttt{Slice\_thickness}$ & 30\\  
$\texttt{method}$ & Lsq3MLs\\  
$\texttt{Niter}$ & 25\\  
$\texttt{delay\_probe\_update}$ & 1\\ 
$\texttt{delay\_drift\_correct}$ & 5\\ 
$\texttt{delay\_dz\_update}$ & 999\\ 
$\texttt{beta\_object}$ & 1\\  
$\texttt{beta\_probe}$ & 1\\
$\texttt{apply\_subpix\_shift}$ & true\\
$\texttt{variable\_intensity}$ & false\\
$\texttt{apply\_multimodal\_update}$ & false\\
object initialization & constant phase\\
probe initialization & defocus from interactive \\
&single-sideband reconstruction \cite{Pelz_2021}\\
\end{tabular}
\end{table}

\newpage
\pagebreak
\section*{Experimental parameters for TEAM 0.5}
\begin{table}[ht!]
\begin{tabular}{ l l}
 Microscope Voltage & \SI{200}{\kilo\eV} \\ 
 Electron gun & S-FEG\\  
 Source size (FWHM) & \SI{0.8}{\angstrom}\\  
 Cc & \SI{0.6}{\milli\meter}\\  
 Defocus spread (FWHM) & \SI{6}{\nano\meter}\\ 
 Convergence semi-angle & \SI{22.3}{\milli\radian}\\  
 Depth of field & \SI{5}{\nano\meter}\\  
 Detector & 4D Camera @ \SI{87}{\kilo\hertz}   \\
 Detector pixel size & \SI{4.14e-3}{\angstrom^{-1}} / \SI{173.6}{\micro\radian}    \\
 Detector pixel size (binned) & \SI{0.0272}{\angstrom^{-1}} / \SI{1.136}{\milli\radian}   \\
 Detector outer angle & \SI{44.6}{\milli\radian}    \\
 Reconstruction pixel size & 0.28 Å\\
 Energy filter & No    \\
 Number of projections & 36    \\
 Tilt range & \SI{-62.2}{\deg}\\
 & \SI{61.7}{\deg}\\
 Total recorded diffraction patterns & \num{23040000}    \\
 STEM step size & \SI{0.33}{\angstrom}    \\
 STEM dwell time & \SI{11.49}{\micro\second}    \\
 Probe current & \SI{70}{\pico\ampere}    \\
 Electron fluence accumulated & \SI{1.6e6}{\elementarycharge\per\angstrom^2}    \\
 Electron fluence per projection & \SI{4.6e4}{\elementarycharge\per\angstrom^2}    \\
 Avg. electrons per diffraction pattern & \num{7376}\\
 \end{tabular}
\caption{\label{tab:data_collection} Experimental parameters for multi-slice ptychographic tomography.}
\end{table}
\section*{Resolution}
\begin{figure*}[ht!]
    \includegraphics[width=1\textwidth]{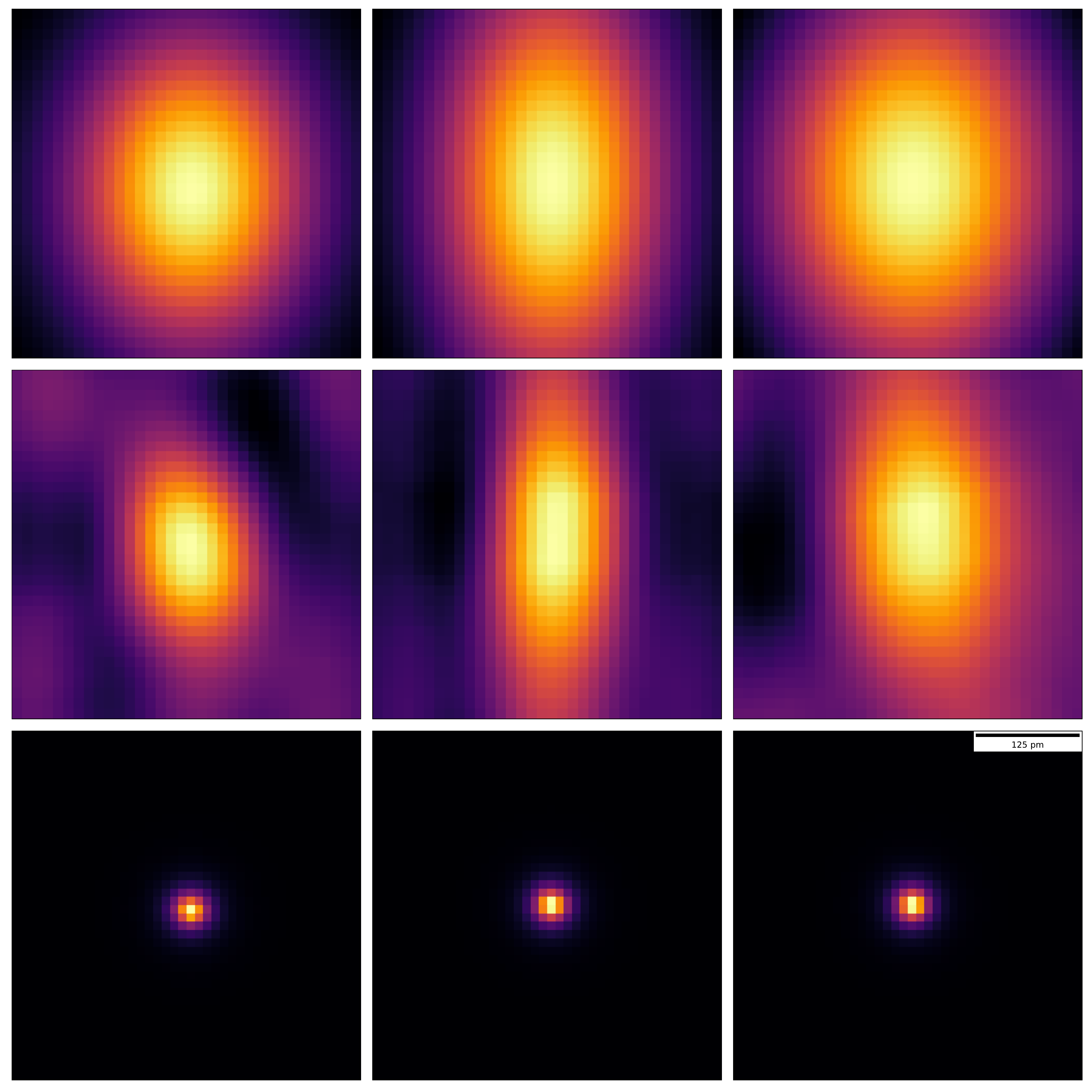}
    \caption{\label{fig:Zr_resolution} Left column: central slice through the Co atomic volume, perpendicular to the missing wedge (X-Y plane). Middle column: central slice through the atomic volume (X-Z plane). Right column: central slice through the atomic volume (Y-Z plane). a)-c) Simulated potential in g)-i), convolved with the error-minimizing Gaussian. d)-f) Mean experimental Co atom. g)-i) Simulated potential, with Debye-Waller blur from 250 Frozen phonons.}
\end{figure*}
\newpage
\pagebreak
\end{document}